\documentclass[aps,prc,preprint,amsmath,showpacs,superscriptaddress,nofootinbib]{revtex4}
\usepackage{graphicx,epsfig,epstopdf}
\newcommand{\beqy}{\begin{eqnarray}}
\newcommand{\eeqy}{\end{eqnarray}}
\newcommand{\bmlet}{\begin{subequations}}
\newcommand{\emlet}{\end{subequations}}

\begin{document}

\textwidth 16.2 cm
\oddsidemargin -.54 cm
\evensidemargin -.54 cm

\def\gsimeq{\,\,\raise0.14em\hbox{$>$}\kern-0.76em\lower0.28em\hbox  
{$\sim$}\,\,}  
\def\lsimeq{\,\,\raise0.14em\hbox{$<$}\kern-0.76em\lower0.28em\hbox  
{$\sim$}\,\,}

\title{Self-interaction errors in nuclear energy density functionals}
\author{N.~Chamel}
\affiliation{Institut d'Astronomie et d'Astrophysique, CP-186, Universit\'e
Libre de Bruxelles, 1050 Brussels, Belgium}

\date{\today}

\begin{abstract}
When applied to a single nucleon, nuclear energy density functionals may 
yield a non-vanishing internal energy thus implying that the nucleon is 
interacting with itself. It is shown how to avoid this unphysical feature 
for semi-local phenomenological functionals containing all possible bilinear 
combinations of local densities and currents up to second order in the derivatives. 
The method outlined in this Rapid Communication could be easily extended to functionals containing 
higher order terms, and could serve as a guide for constraining the time-odd part of 
the functional.
\end{abstract}

\pacs{17.30.Fe, 17.60.Jz, 71.15.Mb}

\maketitle

\section{Introduction}

The density functional theory (DFT)~\cite{ko99} has been very successfully employed in a wide variety of fields, 
from chemistry~\cite{koc01} to condensed matter physics~\cite{mar04,bul07,rup09}. 
A somehow similar approach called the nuclear energy density functional (EDF) theory, has been developed in nuclear physics~\cite{drut10}. 
Even though the EDF theory is very often referred to as nuclear DFT, it is conceptually different due to the breaking 
of various symmetries in its current formulations (see e.g. ~\cite{eng07,mes09,dug10}). The EDF theory is a method of choice 
for describing the structure and the dynamics of medium-mass and heavy nuclei. Phenomenological 
nuclear functionals are now able to fit essentially all experimental nuclear mass data with rms deviations falling below 
0.6 MeV~\cite{gcp09}. Moreover the EDF theory is particularly well-suited for studying the dense inhomogeneous nuclear matter 
found in neutron stars and supernova cores. 

Nevertheless it has been known for a long time in condensed matter physics that the internal energy 
of a single electron as calculated by the DFT may be non-zero owing to a spurious interaction of the electron with itself 
(for a recent review, see for instance Section I.B of Ref.~\cite{kum08} and references therein). This problem is not inherent 
to the DFT, but is induced by the use of approximate expressions for the electron exchange-correlation part which violate 
the Pauli exclusion principle. The presence of self-interaction causes various errors and is currently one of the main deficiencies 
of existing density functionals. For instance, this leads to an artificial spreading of the electron density owing to an 
incorrect asymptotic behavior of the Kohn-Sham potential. This spurious effect spoils the calculations of chemical properties. 
In particular, functionals contaminated by self-interactions fail to correctly describe the dissociation behavior of some homonuclear 
and heteronuclear diatomic radicals. The removal of self-interactions in the DFT has lead to the development of orbital-dependent 
density functionals~\cite{per81,kum08}. 

This self-interaction problem was also recognized a long time ago in nuclear physics~\cite{rip69,str78}, but it has been barely 
discussed in the literature until recently~\cite{bhr03,lac09,ben09,dug09,dug10,drut10,erl10}. Nuclear functionals have been traditionally 
derived from effective interactions using the Hartree-Fock approximation~\cite{bhr03, sto07} which guarantees that in the 
one-particle limit, the internal energy correctly vanishes due to the exact cancellation between direct and exchange terms.
On the other hand, the development of nuclear functionals which are not 
directly constructed from effective forces~\cite{fay00,car08,bal10} calls for a closer examination of their internal consistency. 

In this Rapid Communication, it will be shown how to eliminate self-interaction errors in the one-nucleon limit in semi-local 
nuclear energy density functionals containing all possible bilinear combinations of local densities, 
currents and their derivatives up to second order.

\section{Skyrme force versus nuclear energy density functional}
\label{sec2}

Throughout this paper, pure nucleon states will be assumed. The more general formalism involving 
neutron-proton mixing has been developed in Ref.~\cite{perl04}. 
The nuclear energy density functionals that we consider here are of the form
\beqy
\label{1}
E=E_{\rm kin}+E_{\rm Coul}+E_{\rm Sky}\quad ,
\eeqy
where $E_{\rm kin}$ is the kinetic energy, $E_{\rm Coul}$ is the Coulomb energy and 
$E_{\rm Sky}=\int{\rm d}^3\pmb{r}\,\mathcal{E}_{\rm Sky}(\pmb{r})$ is the nuclear energy. 
The latter is a functional of the density matrix in coordinate space assumed to be 
purely diagonal in isospin space with elements $\rho_q(\pmb{r}, \sigma; \pmb{r^\prime}, \sigma^\prime)$ 
with $q=n,p$ for neutron and proton respectively (denoting the spin states by $\sigma,\sigma^\prime=1,-1$ 
for spin up and spin down respectively). In particular, $\mathcal{E}_{\rm Sky}(\pmb{r})$ 
can be expressed in terms of the following 
local densities and currents ($\pmb{\hat\sigma}$ is used to indicate the Pauli spin matrices and 
$\mu$, $\nu$, $\kappa$ are indices of Cartesian tensors): 

\noindent (i) the density
\beqy
\label{2}
\rho_q(\pmb{r}) = \sum_{\sigma=\pm 1}\rho_q(\pmb{r}, \sigma; \pmb{r}, \sigma)\quad ,
\eeqy
(ii) the kinetic density
\beqy
\label{3}
\tau_q(\pmb{r}) = \sum_{\sigma=\pm 1}\int\,{\rm d}^3\pmb{r^\prime}\,\delta(\pmb{r}-\pmb{r^\prime}) \pmb{\nabla}\cdot\pmb{\nabla^\prime}
\rho_q(\pmb{r}, \sigma; \pmb{r^\prime}, \sigma)\quad ,
\eeqy
(iii) the current-vector density
\beqy
\label{4}
\pmb{j_q}(\pmb{r})=-\frac{\rm i}{2}\sum_{\sigma=\pm 1}\int\,{\rm d}^3\pmb{r^\prime}\,\delta(\pmb{r}-\pmb{r^\prime}) (\pmb{\nabla} -\pmb{\nabla^\prime})\rho_q(\pmb{r}, \sigma; \pmb{r^\prime}, \sigma)\quad ,
\eeqy
(iv) the spin pseudovector density
\beqy
\label{5}
\pmb{s_q}(\pmb{r})= \sum_{\sigma,\sigma^\prime=\pm 1}\rho_q(\pmb{r}, \sigma; \pmb{r}, \sigma^\prime)\langle\sigma^\prime|\pmb{\hat\sigma}|\sigma\rangle\quad ,
\eeqy
(v) the spin pseudovector kinetic density
\beqy
\label{6}
T_{q \mu}(\pmb{r}) = \sum_{\sigma,\sigma^\prime=\pm 1}\int\,{\rm d}^3\pmb{r^\prime}\,\delta(\pmb{r}-\pmb{r^\prime}) \pmb{\nabla}\cdot\pmb{\nabla^\prime}
\rho_q(\pmb{r}, \sigma; \pmb{r^\prime}, \sigma^\prime)\langle\sigma^\prime|\hat\sigma_\mu|\sigma\rangle\quad ,
\eeqy
(vi) the spin current pseudotensor density
\beqy
\label{7}
J_{q \mu\nu}(\pmb{r})=-\frac{\rm i}{2}\sum_{\sigma,\sigma^\prime=\pm 1}\int\,{\rm d}^3\pmb{r^\prime}\,\delta(\pmb{r}-\pmb{r^\prime}) (\nabla_\mu -\nabla_\mu^\prime)\rho_q(\pmb{r}, \sigma; \pmb{r^\prime}, \sigma^\prime)\langle\sigma^\prime|\hat\sigma_\nu|\sigma\rangle \quad ,
\eeqy
and (vii) the tensor-kinetic pseudovector density 
\beqy
\label{8}
F_{q \mu}=\frac{1}{2}\sum_{\sigma,\sigma^\prime=\pm 1}\sum_\nu\int\,{\rm d}^3\pmb{r^\prime}\,\delta(\pmb{r}-\pmb{r^\prime}) (\nabla_\mu\nabla^\prime_\nu+\nabla^\prime_\mu\nabla_\nu)\rho_q(\pmb{r}, \sigma; \pmb{r^\prime}, \sigma^\prime)\langle\sigma^\prime|\hat\sigma^\nu|\sigma\rangle \quad .
\eeqy 

It is convenient to introduce the isospin index $t=0,1$ for isoscalar and isovector quantities respectively.
Isoscalar quantities (also written without any subscript) are sums over neutrons and protons (e.g. 
$\rho_0=\rho=\rho_n+\rho_p$) while isovector quantities are differences between neutrons and protons 
(e.g. $\rho_1=\rho_n-\rho_p$). 
The energy density $\mathcal{E}_{\rm Sky}$ constructed from all possible bilinear terms
up to second order in the derivatives that are invariant under time reversal, space 
inversions and rotations, is given by
\bmlet 
\beqy
\label{9a}
\mathcal{E}_{\rm Sky}=\sum_{t=0,1}(\mathcal{E}_t^{\rm even}+\mathcal{E}_t^{\rm odd})\quad, 
\eeqy
\beqy
\label{9b}
\mathcal{E}_t^{\rm even}&=&C_t^\rho \rho_t^2+C_t^{\Delta\rho}\rho_t\Delta\rho_t+C_t^\tau\rho_t\tau_t
+C_t^{\nabla J}\rho_t\nabla\cdot\pmb{J_t}+C_t^J \sum_{\mu,\nu}J_{t,\mu\nu}J_{t,\mu\nu}\nonumber\\
&+&\frac{1}{2}C_t^{TrJ}\left(\sum_{\mu} J_{t,\mu\mu}\right)^2
+\frac{1}{2}C_t^{J^2}\sum_{\mu,\nu} J_{t,\mu\nu} J_{t,\nu\mu}\quad ,
\eeqy
\beqy
\label{9c}
\mathcal{E}_t^{\rm odd}&=&C_t^s s_t^2+C_t^{\Delta s}\pmb{s_t}\cdot \Delta \pmb{s_t}+C_t^T \pmb{s_t}\cdot\pmb{T_t}
+C_t^j j_t^2+C_t^{\nabla j} \pmb{s_t}\cdot\nabla\times\pmb{j_t}+C_t^{\nabla s}(\pmb{\nabla}\cdot\pmb{s_t})^2 \nonumber\\
&+&C_t^F \pmb{s_t}\cdot\pmb{F_t}\quad .
\eeqy
\emlet 
The ``time-even'' part $\mathcal{E}_t^{\rm even}$ (``time-odd'' part $\mathcal{E}_t^{\rm odd}$) contains only even (odd) 
densities and currents with respect to time reversal. 
In situations for which the time-reversal symmetry is preserved, the time-odd densities $\pmb{j_q}(\pmb{r})$, 
$\pmb{s_q}(\pmb{r})$, $\pmb{T_q}(\pmb{r})$ and $\pmb{F_q}(\pmb{r})$ must vanish. 
The coupling ``constants'' $C_t^\rho$ and $C_t^s$ generally depend on the isoscalar density $\rho=\rho_n+\rho_p$
as follows
\beqy
\label{10}
C_t^\rho=a_t^\rho + b_t^\rho \rho^\alpha\quad ,
\eeqy
\beqy
\label{11}
C_t^s=a_t^s + b_t^s \rho^\alpha\quad .
\eeqy
The spin-current vector density $\pmb{J_q}(\pmb{r})$ appearing in Eq.~(\ref{9b}) is defined by
\beqy
\label{12}
J_{q \mu}(\pmb{r}) = \sum_{\nu\kappa}\epsilon_{\mu\nu\kappa} J_{q \nu\kappa}
\eeqy
where $\epsilon_{\mu\nu\kappa}$ is the Levi-Civita tensor. 
Gauge (including Galilean) invariance imposes the 
following relations~\cite{eng75,doba95}
\beqy
\label{13}
C_t^j=-C_t^\tau\, , \quad C_t^J=-C_t^T\, ,\quad C_t^{\nabla j}=C_t^{\nabla J}\, , \quad C_t^{TrJ}=-C_t^F=C_t^{J^2}\quad .
\eeqy

Historically the type of functionals given by Eqs.~(\ref{1}) and (\ref{9a})--(\ref{9c}) 
was obtained from the Hartree-Fock approximation using effective zero-range interactions 
of the Skyrme type~\cite{bhr03,perl04,sto07,les07}
\beqy
\label{14}
v_{i,j} & = & 
t_0(1+x_0 P_\sigma)\delta({\pmb{r}_{ij}})
+\frac{1}{2} t_1(1+x_1 P_\sigma)\frac{1}{\hbar^2}\left[p_{ij}^2\,
\delta({\pmb{r}_{ij}}) +\delta({\pmb{r}_{ij}})\, p_{ij}^2 \right]\nonumber\\
& &+t_2(1+x_2 P_\sigma)\frac{1}{\hbar^2}\pmb{p}_{ij}\cdot\delta(\pmb{r}_{ij})\,
 \pmb{p}_{ij}
+\frac{1}{6}t_3(1+x_3 P_\sigma)\rho(\pmb{r})^\alpha\,\delta(\pmb{r}_{ij})
\nonumber\\
& &+\frac{\rm i}{\hbar^2}W_0(\pmb{\hat\sigma_i}+\pmb{\hat\sigma_j})\cdot
\pmb{p}_{ij}\times\delta(\pmb{r}_{ij})\,\pmb{p}_{ij}   \nonumber\\
& &+\frac{1}{2}t_e\biggl\{\bigl[3(\pmb{\hat\sigma_i}\cdot\pmb{p}_{ij})(\pmb{\hat\sigma_j}\cdot\pmb{p}_{ij}) 
-(\pmb{\hat\sigma_i}\cdot\pmb{\hat\sigma_j})p_{ij}^2\bigr]\delta(\pmb{r}_{ij})
+\delta(\pmb{r}_{ij})\bigl[3(\pmb{\hat\sigma_i}\cdot\pmb{p}_{ij})(\pmb{\hat\sigma_j}\cdot\pmb{p}_{ij}) 
-(\pmb{\hat\sigma_i}\cdot\pmb{\hat\sigma_j})p_{ij}^2\bigr]\biggr\}\nonumber\\
& &+t_o \bigl[3(\pmb{\hat\sigma_i}\cdot\pmb{p}_{ij})\delta(\pmb{r}_{ij})(\pmb{\hat\sigma_j}\cdot\pmb{p}_{ij}) 
-(\pmb{\hat\sigma_i}\cdot\pmb{\hat\sigma_j})\pmb{p}_{ij}\cdot\delta(\pmb{r}_{ij})\pmb{p}_{ij}\bigr]
\quad ,
\eeqy
where $\pmb{r}_{ij} = \pmb{r}_i - \pmb{r}_j$, $\pmb{r} = (\pmb{r}_i + 
\pmb{r}_j)/2$, $\pmb{p}_{ij} = - {\rm i}\hbar(\pmb{\nabla}_i-\pmb{\nabla}_j)/2$
is the relative momentum and $P_\sigma$ is the two-body spin-exchange 
operator. The relations between the coupling constants in Eqs.~(\ref{9b}) and (\ref{9c}) 
and the parameters of the effective force in Eq.~(\ref{14}), can be found 
for instance in Table I of Ref.~\cite{perl04}. 
The parameters of the force are generally fitted to a selected set of nuclear data, mainly involving 
the time-even part of the functional.
The coupling constants of the time-odd terms are then obtained from those of the 
time-even terms making use of the forementioned correlations (see e.g. Tables IV and VI in Ref.~\cite{doba95}). 
However, since time-odd terms are not directly fitted to experimental data it is rather unlikely that 
nuclear properties involving those terms will be correctly described in this way. In particular, one 
of the main deficiencies of Skyrme forces is that they predict the occurence of spurious transitions 
to spin-ordered phases in the dense matter found in neutron stars and supernova cores~\cite{cgp09, cao10,cg10}. 
It has been recently found that finite-size instabilities could arise at even lower densities thus spoiling
self-consistent calculations in finite nuclei~\cite{les06,kl10}.

Over the past decades, the conventional wisdom has been to regard the nuclear energy density functional
as being more fundamental than effective forces and the development of microscopic functionals using 
many-body techniques with realistic nucleon-nucleon interactions has triggered a burst of activity~\cite{drut10}. 
Unfortunately such kinds of \textit{ab initio} nuclear energy density functionals able to reproduce existing 
experimental nuclear data with the same degree of accuracy as effective interactions are not yet available. 
In the meantime semi-microscopic functionals have been constructed by first fitting the bulk part of 
the functional to realistic calculations of uniform infinite nuclear matter, and second adding a phenomenological surface 
part whose parameters are adjusted to reproduce properties of finite nuclei~\cite{lns, bal10}. 
A complementary approach guided by effective field theories consists in constructing semi-local phenomenological functionals 
from combinations of local densities and currents up to a given order in the derivatives that are invariant under time reversal, 
space inversion, spatial rotations and gauge transformations~\cite{car08}. The expansion in terms of higher-order 
derivatives has been shown to converge rapidly~\cite{car10}. The coupling constants associated with each term are then 
directly fitted to a selected set of nuclear data~\cite{kort10}. However, such kinds of unrestricted fits could lead to formal 
inconsistencies. In particular the resulting functional may allow a nucleon to interact with itself. As will be discussed in the 
next section, this unphysical feature can be avoided by imposing suitable constraints on the coupling constants of the functional.

\section{Removal of self-interaction errors}

The cancellation of spurious self-interactions in nuclear energy density functionals was studied in Ref.~\cite{str78}. 
The main assumption was that the nuclear functional can be calculated from the expectation value of some general Hamiltonian 
with a Slater determinant. However it remains to be proved that the exact functional can be obtained in this way. Moreover 
the nuclear functional was supposed to be purely local and depend on the densities~(\ref{2}) and the spin densities~(\ref{5}) 
only. In the following, the self-interaction problem will be reconsidered for any semi-local functionals
given by Eqs.~(\ref{1}) and (\ref{9a})--(\ref{9c}). 

In the one-particle limit, the density matrix is simply given by 
$\rho_q(\pmb{r}, \sigma; \pmb{r^\prime}, \sigma^\prime)=\varphi^{(q)}(\pmb{r^\prime}\sigma^\prime)^{*}\,\varphi^{(q)}(\pmb{r}\sigma)$
where $\varphi^{(q)}(\pmb{r}\sigma)$ is the single-nucleon wavefunction and the asterisk indicates complex conjugation. 
Inserting this expression into Eqs.~(\ref{2})--(\ref{8}), substituting in 
Eqs.~(\ref{9a})--(\ref{9c}) and making use of the gauge invariance Eq.~(\ref{13}) yield

\beqy
\label{15}
\mathcal{E}^{\rm even}_t+\mathcal{E}^{\rm odd}_t&=&(C_t^\rho+C_t^s)\varphi^\dagger \varphi
+\biggl[\frac{1}{4}( C_t^\tau+C_t^T)-(C_t^{\Delta\rho}+C_t^{\Delta s})\biggr]\biggl[ \varphi^\dagger \nabla \varphi + (\nabla \varphi^\dagger)\varphi \biggr]^2 \nonumber\\
&+&(C_t^\tau-2C_t^T-4C_t^{\Delta s}-C_t^F) \biggl[ (\varphi^\dagger\varphi)(\nabla\varphi^\dagger)\cdot \nabla\varphi-\varphi^\dagger(\nabla\varphi)\cdot(\nabla\varphi^\dagger)\varphi\biggr] \nonumber\\
&+&\biggl(\frac{1}{4} C_t^F+C_t^{\nabla s}\biggr)\biggl[\sum_\mu\varphi^\dagger\hat\sigma^\mu\nabla_\mu\varphi+ (\nabla_\mu\varphi^\dagger)\hat\sigma^\mu\varphi\biggr]^2
\eeqy
where $\varphi$ denotes the two-component spinor wavefunction and the superscript $\dagger$ indicates
the adjoint operation. 
Now the functional should be devoid of self-interactions whether the nucleon is free or is embedded in an 
external potential, i.e. for \emph{any} single-nucleon wavefunction. The requirement that the nuclear energy 
$E_{\rm Sky}$ vanishes thus leads to the four identities
\beqy
\label{16}
C_0^\rho+C_1^{\rho}+C_0^s+C_1^s=0\quad ,
\eeqy
\beqy
\label{17}
C_0^\tau+C_1^{\tau}+C_0^T+C_1^T=4(C_0^{\Delta\rho}+C_1^{\Delta\rho}+C_0^{\Delta s}+C_1^{\Delta s})\quad ,
\eeqy
\beqy
\label{18}
4(C_0^{\nabla s}+C_1^{\nabla s})+C_0^F+C_1^F=0  \quad ,
\eeqy
\beqy
\label{19}
C_0^\tau+C_1^\tau-2(C_0^T+C_1^T)-(C_0^F+C_1^F)-4(C_0^{\Delta s}+C_1^{\Delta s})=0  \quad .
\eeqy

If the coefficients $C_t^\rho$ and $C_t^s$ are allowed to depend on density according to 
Eqs.~(\ref{10}) and (\ref{11}), following the same analysis entails that Eq.~(\ref{16}) 
would have to be replaced by 
\bmlet
\beqy
\label{20a}
a_0^\rho+a_1^{\rho}+a_0^s+a_1^s=0\quad ,
\eeqy
\beqy
\label{20b}
b_0^\rho+b_1^{\rho}+b_0^s+b_1^s=0\quad .
\eeqy
\emlet 
Note that these two conditions lead to Eq.~(\ref{16}), but the reverse is not necessarily true. 
Equation~(\ref{16}), was already noticed in Ref.~\cite{bend02}, even though its origin was not discussed. 

Equations~(\ref{16})--(\ref{19}) can be automatically satisfied by first 
fitting the parameters of a Skyrme force given by Eq.~(\ref{14}) and then calculating the 
corresponding coupling constants. But this usual fitting protocol imposes a lot more relations 
on the different terms of the functional than what is required by the cancellation of self-interactions 
(see e.g. Tables IV and VI in Ref.~\cite{doba95}). Actually these relations are generally 
not strictly enforced so that the functional may be contaminated by self-interactions. For example, 
the coupling constant $C_t^J$ (hence also $C_t^T$ due to gauge invariance) is set to zero in most 
Skyrme parametrizations~\cite{cg10}. It has been suggested to also drop the isoscalar terms associated with 
$C_0^s$ and $C_0^{\Delta s}$ in the time-odd part of the functional~\cite{sto07,klup09}. 

Self-interaction errors in the one-particle limit can contaminate systems consisting of many particles. 
Let us consider for instance cold fully polarized neutron matter. Assuming that all spins are aligned along the 
$z$-axis and neglecting the anisotropies induced by the polarization, Eqs.~(\ref{9a})--(\ref{9c}) lead to 
the following expression for the energy density of polarized neutron matter
\beqy
\label{21}
\mathcal{E}^{\rm pol}_{\rm NeuM}&=&\left[\frac{\hbar^2}{2M_n}+\biggl(C_0^\tau+C_1^\tau+C_0^T+C_1^T+\frac{1}{3}(C_0^F+C_1^F)\biggr)\rho\right]\tau^{\rm pol}_{n}\nonumber\\
& & +(C_0^\rho+C_1^\rho+C_0^s+C_1^s)\rho^2\quad ,
\eeqy
with the kinetic density
\beqy
\label{22}
\tau^{\rm pol}_{n}=\frac{3}{5}(6\pi^2)^{2/3}\rho^{5/3}\quad .
\eeqy
It can thus be seen from Eqs.~(\ref{16}) and (\ref{21}) that the 
error caused by self-interactions is given by 
\beqy
\label{23}
\delta \mathcal{E}^{\rm pol}_{\rm NeuM}=(C_0^\rho+C_1^\rho+C_0^s+C_1^s)\rho^2\quad .
\eeqy
As expected the magnitude of this spurious term grows with increasing density and could thus lead to large 
deviations when such functionals are applied to the high-density matter found in neutron stars. 
In particular if $\delta \mathcal{E}^{\rm pol}_{\rm NeuM}<0$, self-interaction errors will drive a 
ferromagnetic collapse of neutron matter with the energy~(\ref{21}) 
decreasing asymptotically with increasing density. Such a pathological behavior is obviously contradicted 
by neutron-star observations. 

\section{Conclusions}

Phenomenological nuclear functionals based on Skyrme forces have been very successful in 
describing various nuclear systems, from finite nuclei to neutron stars and supernova cores. 
Nevertheless the use of effective forces introduces tight correlations between different terms 
of the functional which hamper further improvements. For instance, adding 
density- and momentum-dependent terms in $t_1$ and $t_2$ contributes to the eight coupling constants 
$C_t^\tau$, $C_t^T$, $C_t^{\Delta\rho}$ and $C_t^{\Delta s}$, and also generates many new terms~\cite{cgp09}. 
Improving one part of the functional may therefore deteriorate other parts. 

A different approach to improve Skyrme functionals systematically consists in constructing semi-local 
functionals from all possible combinations of local densities and currents up to a given order in the 
derivatives, that are invariant under time reversal, space inversion, spatial rotations and gauge 
transformations~\cite{car08}. The coupling constants appearing in the functional are treated as free 
adjustable parameters and are directlty fitted to a selected set of nuclear data. Unfortunately such 
functionals inspired by effective field theories may allow a single nucleon to interact with itself thus 
indicating a violation of the Pauli exclusion principle. 
This unphysical feature can be avoided by imposing suitable restrictions on the coupling constants. 
These constraints have been derived for semi-local functionals up to second order in the derivatives 
and are given by Eqs.~(\ref{16})--(\ref{19}). The method outlined in this Rapid Communication can be easily extended 
to functionals of higher order. 

Self-interaction errors in the one-particle limit can also contaminate systems consisting of many particles. 
For instance, it has been shown that in some cases these errors can lead to a catastrophic ferromagnetic 
collapse of neutron matter in contradiction with neutron-star observations. More generally, because a single 
nucleon breaks time-reversal symmetry, preventing a nucleon from interacting with itself necessarily requires 
the presence of time-odd densities and currents in the functional. The cancellation of self-interactions could 
thus serve as a guide for constraining the time-odd part of the functional, which still remains poorly known. 

Even though Eqs.~(\ref{16})--(\ref{19}) ensure the vanishing of the internal energy in the limit of a single 
nucleon, other kinds of self-interactions and self-pairing could still arise in many-body calculations~\cite{ben09}. 
Nuclear energy density functionals obtained from the Hartree-Fock method with effective forces may also be spoiled 
by such \emph{many-body} self-interactions. In particular, it is well-known that the density-dependent $t_3$ term in the 
Skyrme force~(\ref{14}) is formally inconsistent from the many-body theory point of view. Whereas its contribution to
the ground-state energy vanishes in the one-nucleon limit, this term contains spurious many-body self-interactions which 
can become manifest when dealing with more than one particle.
Correcting for these many-body self-interactions would provide additional constraints on nuclear 
energy density functionals.

{\it Acknowledgments}. 
This work was financially supported by FNRS (Belgium), Communaut\'e fran\c{c}aise de Belgique 
(Actions de Recherche Concert\'ees) and by CompStar (a Research Networking Programme of the European 
Science Foundation). The author is grateful to J.M. Pearson for valuable comments.

\end{document}